\documentclass[sigconf, nonacm]{acmart}

\AtBeginDocument{%
  }

\begin{document}


\title[Pensieve Discuss]{Pensieve Discuss: \\Scalable Small-Group CS Tutoring System with AI}


\author{Yoonseok Yang}
\authornote{Both authors contributed equally to this research.}
\authornote{Corresponding Author}
\orcid{1234-5678-9012}
\affiliation{%
  \institution{Pensieve Inc.}
  \city{San Francisco}
  \state{CA}
  \country{USA}
}
\email{yoon@pensieve.co}

\author{Jack Liu}
\authornotemark[1]
\affiliation{%
  \institution{Pensieve Inc.}
  \city{San Francisco}
  \state{CA}
  \country{USA}}
\email{jack@pensieve.co}

\author{J.D. Zamfirescu-Pereira}
\affiliation{%
  \institution{UC Berkeley}
  \city{Berkeley}
  \state{CA}
  \country{USA}
}
\email{zamfi@berkeley.edu}

\author{John DeNero}
\authornotemark[2]
\affiliation{%
  \institution{UC Berkeley}
  \city{Berkeley}
  \state{CA}
  \country{USA}
}
\email{denero@berkeley.edu}






\begin{abstract}
  Small-group tutoring in Computer Science (CS) is effective, but presents the challenge of providing a dedicated tutor for each group and encouraging collaboration among group members at scale. We present \textit{Pensieve Discuss}, a software platform that integrates synchronous editing for scaffolded programming problems with online human and AI tutors, designed to improve student collaboration and experience during group tutoring sessions. Our semester-long deployment to 800 students in a CS1 course demonstrated consistently high collaboration rates, positive feedback about the AI tutor's helpfulness and correctness, increased satisfaction with the group tutoring experience, and a substantial increase in question volume. The use of our system was preferred over an interface lacking AI tutors and synchronous editing capabilities. Our experiences suggest that small-group tutoring sessions are an important avenue for future research in educational AI.
  
\end{abstract}

\ccsdesc[500]{Social and professional topics~CS1}
\ccsdesc[500]{Applied computing~Computer-assisted instruction}


\keywords{Small-group Tutoring, Artificial Intelligence, Large Language Models, AI Tutors, Collaborative Learning}


\maketitle

\section{Introduction}

Small-group tutoring sessions involve a single human tutor or teaching assistant working with a small group of students to review course material and solve example problems. Research has demonstrated that this mode of instruction is effective for collaborative learning, resistance to forgetting, and instructional efficiency \cite{collaborative-its, smallgroupengaging, smallgroupengaginglab, smallgroupeffective}. Collaborative systems have been shown to significantly enhance learning gains compared to non-collaborative systems in CS education \cite{collaborative-its}.

However, resource constraints can make it impractical to assign a dedicated tutor for each small-group tutoring session in a large course. When a single TA must oversee multiple small groups simultaneously, it can be challenging to monitor each student's progress and provide sufficient support and feedback. The increasing interest in CS courses has exacerbated this issue, making it difficult for educators to maintain high-quality instruction and support during scheduled group tutoring sessions.

Moreover, encouraging collaboration among group members poses another challenge. When each student in a small group is using a computer to write and test code using their own individual editors or writing code on individual worksheets, they may work at different paces and focus their attention on their own individual answers rather than group discussion. Individually distributed web pages or worksheets inhibit students from easily seeing each other's approaches, making collaboration more difficult.


\begin{figure*}[ht!]
  \centering
  \includegraphics[width=0.75\textwidth]{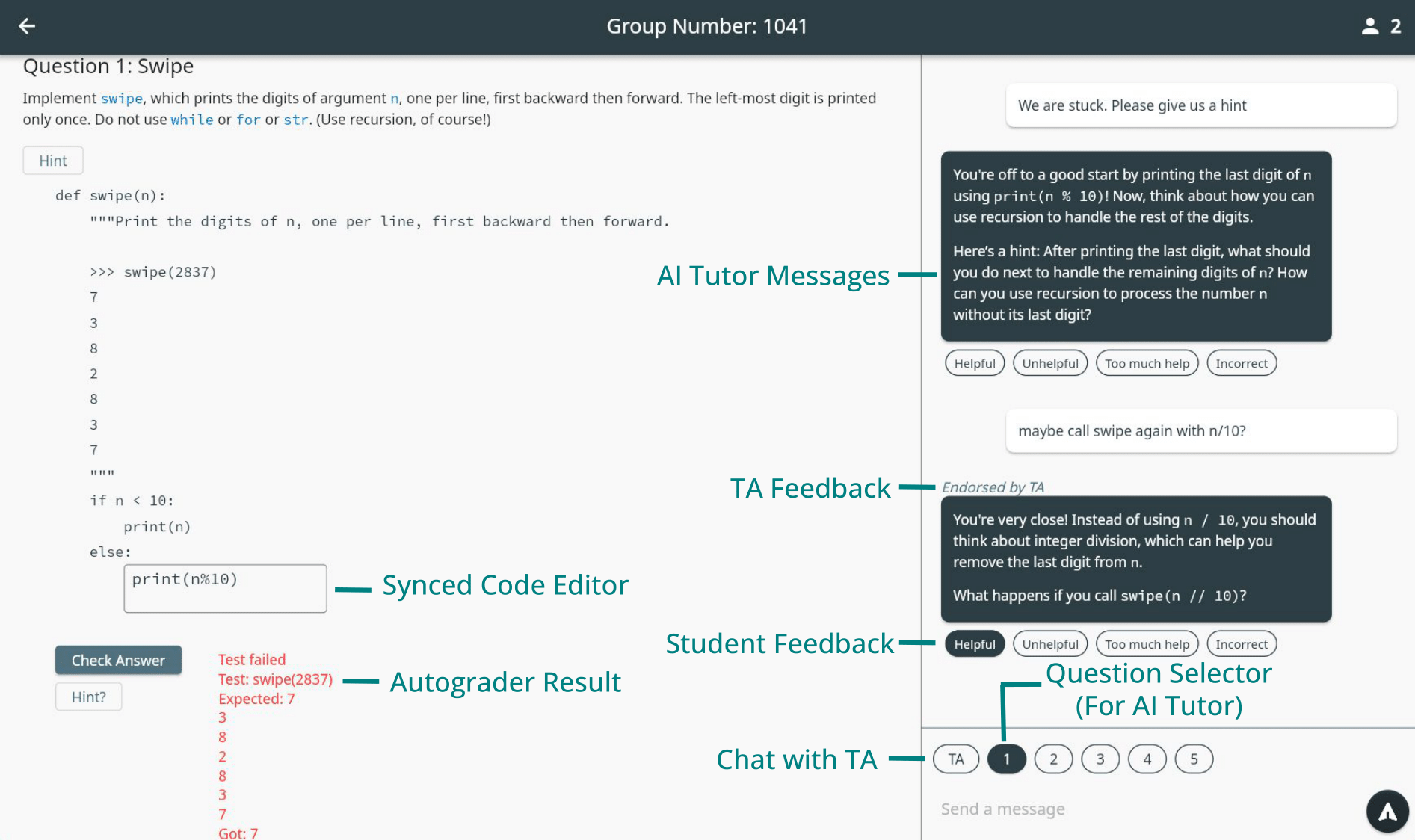}
  \caption{Pensieve Discuss Interface (Student View)}
  \Description{Placeholder}
  \label{fig:teaser}
\end{figure*}

We present \textit{Pensieve Discuss}, a software platform designed to facilitate small-group CS problem solving sessions at scale by combining synchronous editing of scaffolded fill-in-the-blank problems with AI and online human feedback. Students can form small groups in \textit{Pensieve Discuss} and work on problems together using a code editor with shared content. When they are stuck, they can get responsive expert help from an AI Tutor or human TAs. Human TAs can monitor student progress and AI tutor responses in real-time through the platform and assist students as needed.

We deployed \textit{Pensieve Discuss} in a CS1 course that had weekly 80 minute in-person discussion sections. These discussion sections were conducted in the format of small-group tutoring sessions, with groups of up to seven students supervised by a TA who supported multiple groups simultaneously. Prior to introducing \textit{Pensieve Discuss}, students in these groups used individual code editors to write down their answers to questions and check their work, and there was only human TA feedback available rather than both human and AI feedback. \textit{Pensieve Discuss} was deployed to 800 students for the entire Spring 2024 semester. Students had the option of using \textit{Pensieve Discuss} or the individual code editor used previously. 

To our knowledge, \textit{Pensieve Discuss} is the first LLM-based system designed to be used in real-time CS small-group discussions alongside teaching assistants. Our main contributions are as follows:
\begin{itemize}
    \item We present the \textit{Pensieve Discuss} system, which provides collaborative and scalable small-group tutoring for CS1 courses.
    \item We share our findings from student surveys, analysis of usage data, and teaching assistant interviews, which show that our system achieves high satisfaction, encourages collaboration among students, and increased question volume.
\end{itemize}

\textit{Pensieve Discuss} is publicly available on https://discuss.pensieve.co.

\section{Related Work}

\subsection{Small-group tutoring}
Small-group tutoring in CS education is a resource-efficient alternative to one-on-one tutoring. One-on-one tutoring from a human expert is one of the most effective ways for novices to develop robust mental models of programming \cite{one-on-one, codeopticon}. Expert tutors can provide timely, targeted, and proactive help based on a close examination of the student's code, addressing individual learning needs promptly and accurately. However, assigning a TA to each student is impractical in higher education due to the limited availability of TAs. Small-group tutoring presents a promising alternative in terms of scalability. 

Small-group tutoring can also increase student engagement through collaboration. Research has shown that collaborative learning environments facilitated by small-group tutoring can increase student engagement and make the learning experience more enjoyable \cite{smallgroupengaging, smallgroupengaginglab}. Some researchers argue that it is even more effective than one-on-one tutoring due to its instructional efficiency, resistance to forgetting, and promotion of cooperative learning \cite{smallgroupeffective}.

However, tools specifically designed for small-group tutoring have been rarely developed. Researchers have developed tools like Codeopticon \cite{codeopticon} and VizProg \cite{vizprog} to provide real-time, one-to-many tutoring interfaces, enabling instructors to support multiple students more efficiently. However, these tools do not explicitly foster student collaboration, as each student is expected to work independently in a non-synced editor. Moreover, without AI assistants, these tools cannot fully eliminate delays in assisting students, as a single human tutor has limited capacity.


\subsection{LLM-Based Tools in CS Education}

Large introductory CS courses in higher education have increasingly adopted Large Language Models (LLMs) to provide responsive support to students and alleviate instructors' workloads. One prominent application is assisting students in debugging their programming assignments \cite{cs50, 61a-bot, codeaid, codehelp, helprequests, codehelp-deploy, Prather_2023}. LLMs can guide students towards correct solutions with high accuracy, and students find these tools helpful \cite{codeaid, codehelp, codehelp-deploy}. Additional uses of LLMs in CS education include answering students' questions in online course forums \cite{cs50, edbot, course-bot} and generating course materials \cite{sarsa2022automatic, cseduai, code-qg, yuan2023evaluatinginstructiontunedlargelanguage}.

LLM-based assistants are new, and so far as we are aware, no LLM tools have been developed specifically to support small-group tutoring sessions that are supported simultaneously by AI and human tutors or teaching assistants. Compared to the individual use of AI assistants, AI tools for small-group tutoring sessions need to consider collaborative features and human(TA)-in-the-loop components to maximize the learning benefit of each tutoring session.


\begin{figure*}[ht!]
  \includegraphics[width=0.76\textwidth]{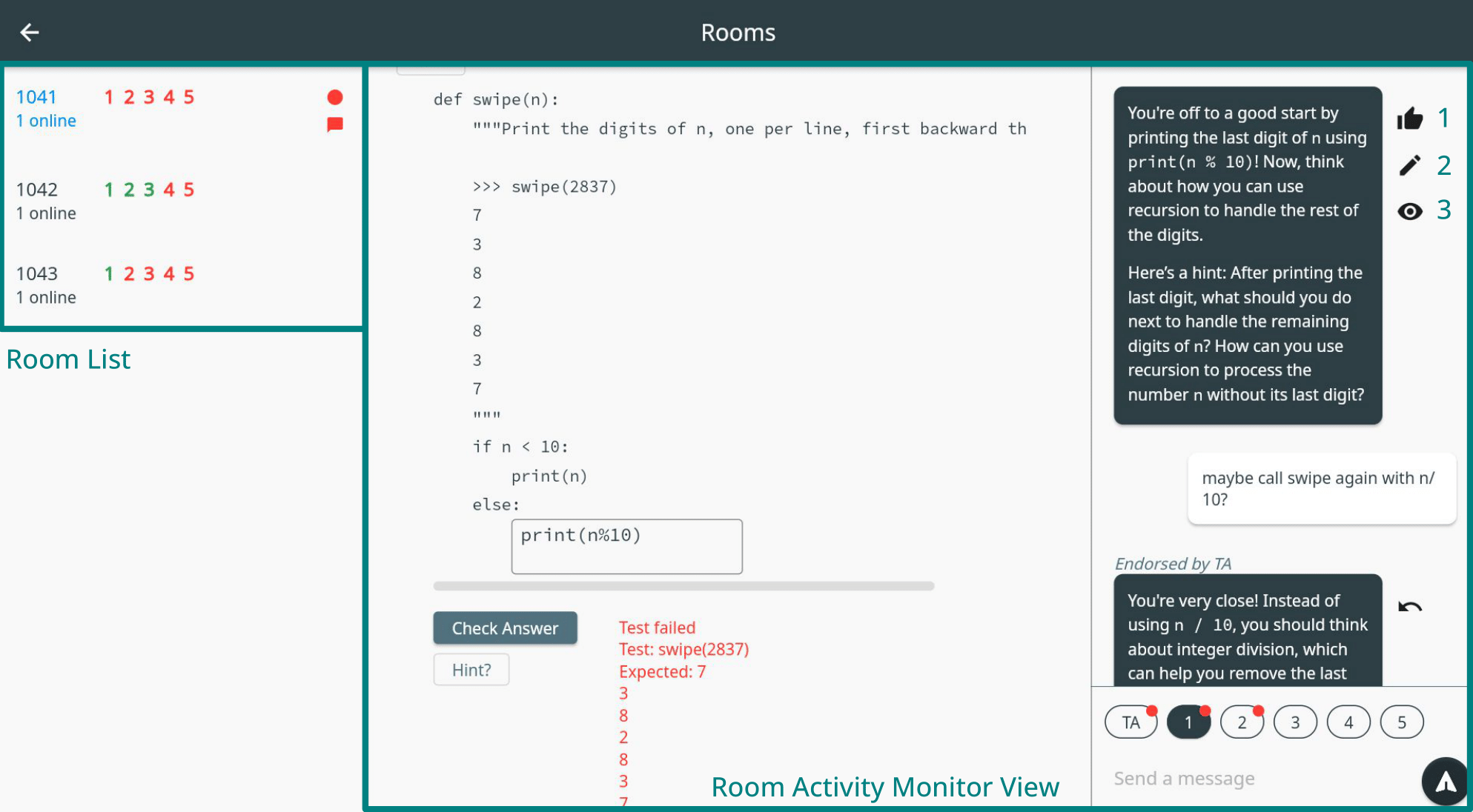}
  \caption{Pensieve Discuss Interface (TA View)}
  \label{fig:ta}
\end{figure*}

\section{System Description}


Our system, \textit{Pensieve Discuss}, was developed to provide an enhanced learning experience to students during real-time in-person small group tutoring sessions that are part of a regular CS1 course offering and are overseen by a TA. Unlike most LLM assistants in CS education intended only for student use, our system serves both students and TAs. In this section, we describe \textit{Pensieve Discuss} in detail for each user group.

\subsection{Student Features}

Figure \ref{fig:teaser} shows the student interface to the system. Students can join a group and solve questions collaboratively. 
 

\textbf{Synced Editor: }Students use a synced editor to write their solutions to coding questions. Similar to Google Docs for programming, this feature allows everyone to see the group's progress and work together towards a solution.


\textbf{Autograder: }When students want to check their answer, they can click "Check Answer" to run a test suite that is created in advance as a component of the problem. Errors during test execution are displayed, enabling students to identify and fix issues in their solutions.


\textbf{AI Tutor: }Any student in the group can interact with the AI Tutor, based on OpenAI's GPT-4 \cite{gpt4}. They can select the question they are currently working on in the Question Selector and type a query for the AI Tutor. The AI Tutor view is synced across group members, so the entire group can see the dialogue between students and the AI Tutor. All students in the group can contribute to the conversation, and the AI Tutor responds to each input. To evaluate the helpfulness of the AI Tutor, we added a feature for students to provide feedback on each AI Tutor message. Students can select one of four feedback labels: \textit{helpful}, \textit{unhelpful}, \textit{too much help}, and \textit{incorrect}.


\textbf{Chat with TA: }Some TAs are physically in the room with students and therefore do not use this interface to communicate with students, but some TAs supervise tutoring groups remotely, and students in sections with a remote TA can talk to their TA through the TA chat view.

\subsection{TA Features}


\textbf{Content Management System (CMS): }TAs can use our CMS to upload questions. The CMS supports markdown, making it easy to import existing content into our platform. TAs can select the programming language and write test cases for the autograder.


\textbf{TA View: }During development, TAs reported that it was very difficult for them to track student progress in any problem-solving session that involved them supervising a large number (such as 30) of students, whether they were in-person or online. To solve this problem, we developed a TA view (Figure \ref{fig:ta}) where TAs can monitor student progress in real-time. They can click each room to observe students' progress, seeing changes in real-time, enabling remote support. When students send a chat to the TA or the AI Tutor, TAs are notified to review new messages. Rooms with unreviewed activity are surfaced to the top of the room list, allowing TAs to prioritize those rooms.


\textbf{TA Feedback: }TAs wanted an easy way to give feedback on AI chat messages. For instance, endorsing an AI message could help students trust the AI's advice. To support this, we developed a TA feedback feature. For each AI message, TAs can provide one of three feedback types: \textit{read}, \textit{endorse}, or \textit{edit}. If a TA endorses a message, students can see that it was endorsed by a TA. TAs can edit a message if it contains incorrect information or gives away the answer to a problem. If TAs don't want to endorse or edit, they can mark the message as "read" to remove the unread message notification.


\subsection{Implementation \& Deployment Details}

When students interact with our AI Tutor, we send a request to GPT-4 containing the question the group is solving, their current solution, and the autograder result if they ran the grader. Each request includes a system prompt that guides GPT-4 on how to assist the group. To prevent GPT-4 from providing solutions, we direct the model to guide them towards the solution by asking questions and offering hints. 



\textit{Pensieve Discuss} was first deployed to 150 students for the final week of the Fall 2023 semester, then to 800 students for the entire Spring 2024 semester. For both semesters, \textit{Pensieve Discuss} was used during 80 minute in-person discussion sections with groups of up to 7 students. Discussion sections were conducted as small-group tutoring sessions, where TAs supervised multiple groups simultaneously. 

For the Spring 2024 semester, six TAs used our system to assist students. Student groups were formed based on scheduling preferences and demographic information to create inclusive study groups \cite{study-group}. Students in each group were assigned a group number at the beginning of the semester, which they used along with their school email to sign into the system. There were 13 weekly discussions throughout the semester, and our system was used for all of them except for the first week.

Among the total 115 groups, 30 groups had an in-person TA, and 85 groups met in person but had a remote TA. A remote TA supported 7-15 groups simultaneously, while an in-person TA supported 5-6 groups at a time. For remote sections, students reached out to their TAs via \textit{Pensieve Discuss} or using Discord\footnote{https://discord.com} voice channels. All student groups collaborated with their peers in person using \textit{Pensieve Discuss}.



\section{Results}
We analyze the data collected during this deployment by focusing on the following Research Questions (RQs):

\indent \textbf{RQ1}: Did our system facilitate student collaboration? 

\indent \textbf{RQ2}: Was the AI Tutor helpful and correct? 

\indent \textbf{RQ3}: Did our system increase student satisfaction?

\indent \textbf{RQ4}: How much did the students and TAs use the chat interface? 

\subsection{Data}

The data presented in this section comes from three sources: 1) student surveys, 2) system usage data, and 3) TA interviews. Students completed short surveys each week directly after completing the discussion section of our CS1 course, in order to receive attendance credit. For the TA interviews, we reached out via email to six TAs who taught discussion sections during the Spring 2024 semester. Two TAs were available for interviews, so we conducted 20-minute virtual interviews with each of them. Both TAs had over two years of experience teaching this CS1 course and used \textit{Pensieve Discuss} to run their remote discussion sections.


\subsection{RQ1: Did our system facilitate student collaboration?}

\begin{figure}[h]
  \centering
  \includegraphics[width=0.85\linewidth]{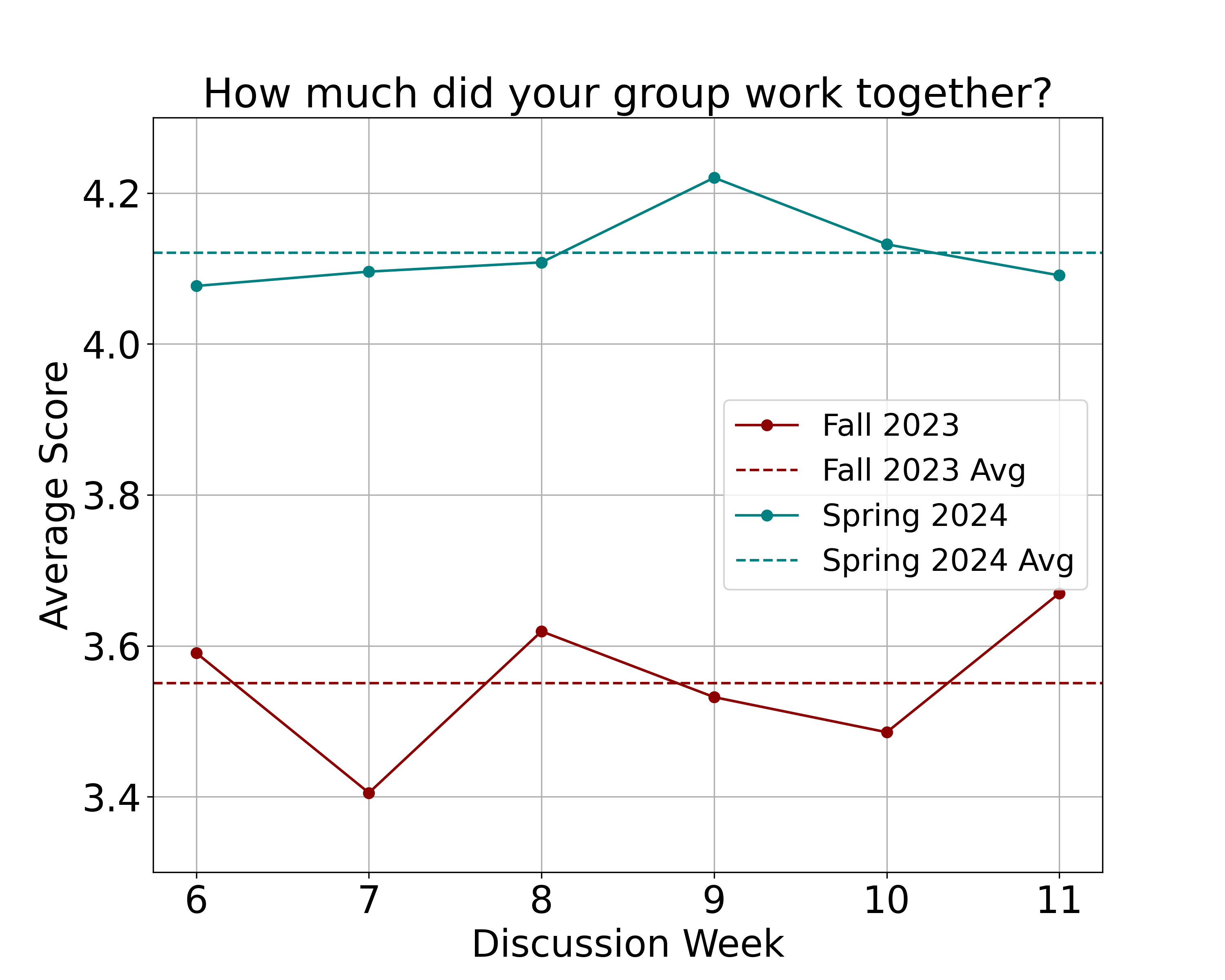}
  \caption{Student collaboration}
  \label{fig:collaboration}
\end{figure}

To assess whether our system facilitated student collaboration, we compared student survey data from the Spring 2024 semester with data from the Fall 2023 semester. \textit{Pensieve Discuss} was not used in the Fall 2023 semester, except for the last one (Discussion 12), but the small-group format overseen by a TA was used in both semesters. We added the survey question about collaboration starting from Discussion 6 in Fall 2023, so we compared data from Discussion 6 to 11 of both semesters; the problems and instructions used in these sessions had largely similar content.

As shown in Figure \ref{fig:collaboration}, student collaboration reported in surveys was consistently higher during the semester when \textit{Pensieve Discuss} was used.

\subsection{RQ2: Was the AI Tutor helpful \& correct?}



\begin{table}[h]
\begin{tabular}{|c|c|}
\hline
\textbf{Very Helpful}     & \textbf{906 (41.8\%)} \\ \hline
Helpful          & 765 (35.3\%)          \\ \hline
Neutral          & 406 (18.8\%)          \\ \hline
Slightly Helpful & 56 (2.6\%)            \\ \hline
Not Helpful      & 32 (1.5\%)            \\ \hline
\end{tabular}
\caption{Aggregate student responses to "How helpful was the AI Tutor?"}
\label{table:helpfulness}
\end{table}

To determine whether the AI Tutor was helpful and correct, we collected student feedback through surveys after each discussion session. As shown in Table \ref{table:helpfulness}, a significant majority of students found the AI Tutor to be helpful, with 41.8\% rating it as \textit{Very Helpful} and 35.3\% as \textit{Helpful}. Only 1.5\% of the responses indicated that the AI Tutor was \textit{Not Helpful}.

Moreover, the AI Tutor's helpfulness was consistently rated high throughout the semester, showing no noticeable trend of decline. This indicates that the AI Tutor maintained a high level of usefulness and accuracy in assisting students with their queries.

\begin{table}[h]
\begin{tabular}{|c|c|}
\hline
\textbf{Helpful}       & \textbf{156} \\ \hline
Unhelpful     & 129          \\ \hline
Too Much Help & 39           \\ \hline
Incorrect     & 42           \\ \hline
\end{tabular}
\caption{Student labeled chat distribution}
\label{table:chat_label_student}
\end{table}

We also analyzed the per-chat feedback labels provided by students for each chat message. Since the functionality to give per-chat feedback was deployed after Discussion 2, 7084 out of 7516 AI messages were able to be labeled. Out of these 7084 messages, only 336 (4.7\%) were labeled by students.

As shown from Figure \ref{table:chat_label_student}, while \textit{Helpful} was the most popular label, the proportion of negative labels (\textit{Unhelpful}, \textit{Too much help}, \textit{Incorrect}) was more substantial compared to the aggregate helpfulness ratings shown in Table \ref{table:helpfulness}. We hypothesize that students are more likely to label negative interactions since helpful chats are considered the default expectation, as reflected in the survey feedback.


TAs corroborated the high accuracy of the AI Tutor during the interview, stating that it was rarely wrong. However, they noted that the AI Tutor sometimes "overcomplicated" explanations, requiring TAs to intervene and simplify the information for students.

\subsection{RQ3: Did our system increase student satisfaction?}

\begin{figure}[h]
  \centering
  \includegraphics[width=0.85\linewidth]{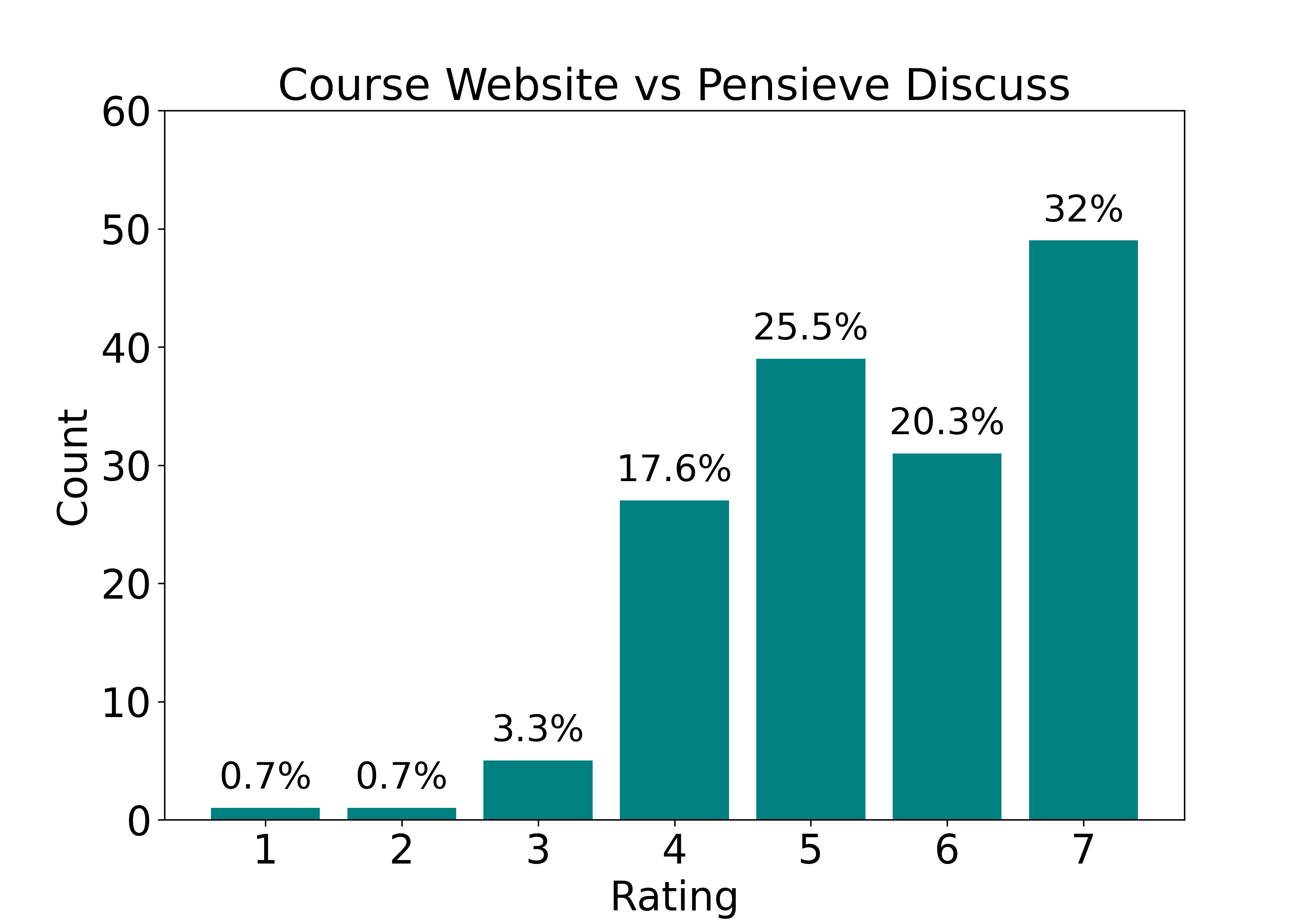}
  \caption{Compared to the course website (individual code editor without AI Tutor), how would you rate Pensieve Discuss?}
  \label{fig:fa23_satisfaction}
\end{figure}

After piloting our system with 150 students during the last discussion section of the Fall 2023 semester, we asked students to compare the previous discussion interface with the new discussion interface using \textit{Pensieve Discuss}. Students rated their preference on a scale of 1 (strongly prefer previous system) to 7 (strongly prefer our system). As shown in Figure \ref{fig:fa23_satisfaction}, 77.8\% of the students preferred our system, with 32\% strongly preferring it over the non-synced individual code editor without AI assistance used previously in Fall 2023.

\begin{figure}[h]
  \centering
  \includegraphics[width=0.85\linewidth]{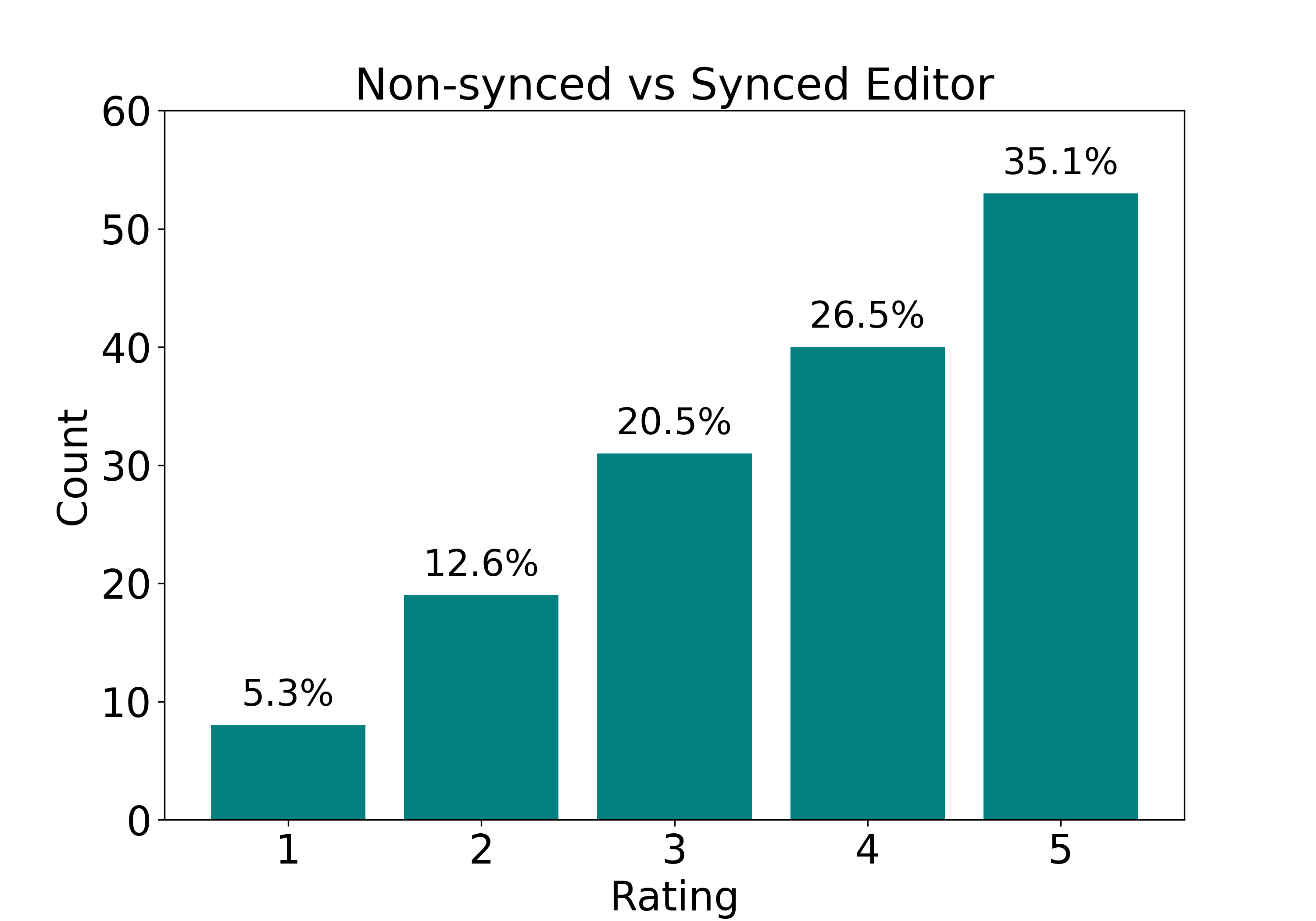}
  \caption{Compared to working in a non-synced editor, did you prefer working with Pensieve Discuss synced editor?}
  \label{fig:fa23_editor}
\end{figure}

We also asked the same students specifically about their experience with the synced editor. They rated their preference on a scale of 1 (strongly prefer non-synced editor) to 7 (strongly prefer synced editor). As shown in Figure \ref{fig:fa23_editor}, the majority of students preferred the synced editor over the non-synced editor, with 35\% strongly preferring the synced editor.

\begin{figure}[h]
  \centering
  \includegraphics[width=0.85\linewidth]{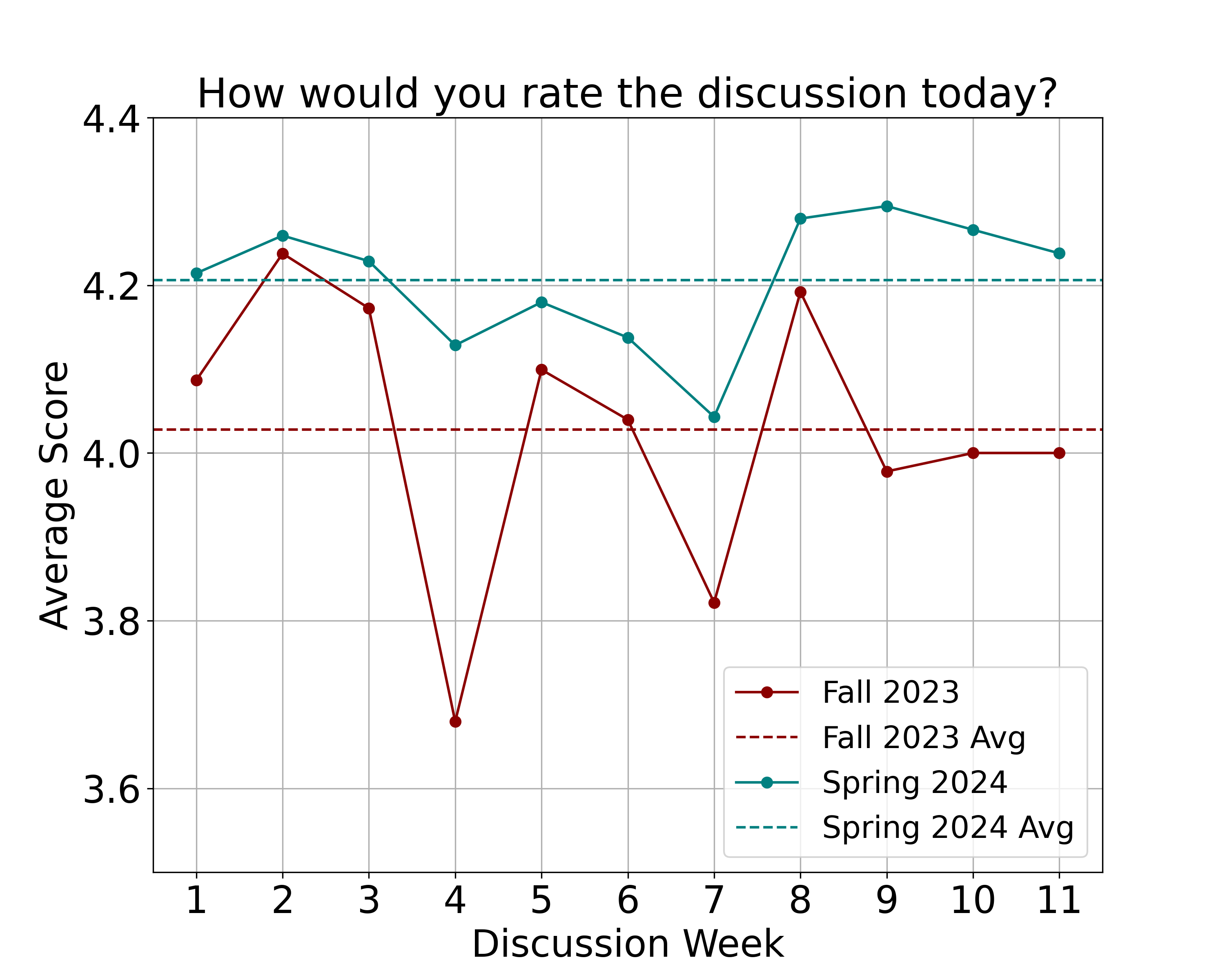}
  \caption{Student satisfaction}
  \label{fig:satisfaction}
\end{figure}

Next, we compare the survey results of Discussions without Pensieve Discuss (Discussion 1 - 11 in Fall 2023 semester) with the survey results of Discussions with Pensieve Discuss (Discussion 1 - 11 in Spring 2024 semester). As shown from Figure \ref{fig:satisfaction}, student satisfaction is consistently higher for the semester (Spring 2024) that used our system. 
We acknowledge that there is a difference in student populations from Fall 2023, which included mostly CS-intended majors, versus Spring 2024 which included students with a much broader range of intended majors.

\begin{table}[h]
\begin{tabular}{|l|l|l|}
\hline
             & AI Helpfulness & Collaboration \\ \hline
Satisfaction & 0.30           & 0.51          \\ \hline
\end{tabular}
\caption{Correlation between AI Helpfulness, Collaboration to Satifaction. }
\label{table:corr}
\end{table}

We examined the correlation between student satisfaction, collaboration, and AI Tutor helpfulness. As shown in Table \ref{table:corr}, student collaboration (r = 0.51) and AI Tutor helpfulness (r = 0.3) were both positively correlated with student satisfaction. This suggests that these factors are important contributors to the student experience during discussion sections.

\subsection{RQ4: How much did students \& TAs use the system?}

In this section, we analyze the interactions between students, TAs, and our platform.

\subsubsection{Student to AI Tutor interaction}

\begin{figure}[h]
  \centering
  \includegraphics[width=0.85\linewidth]{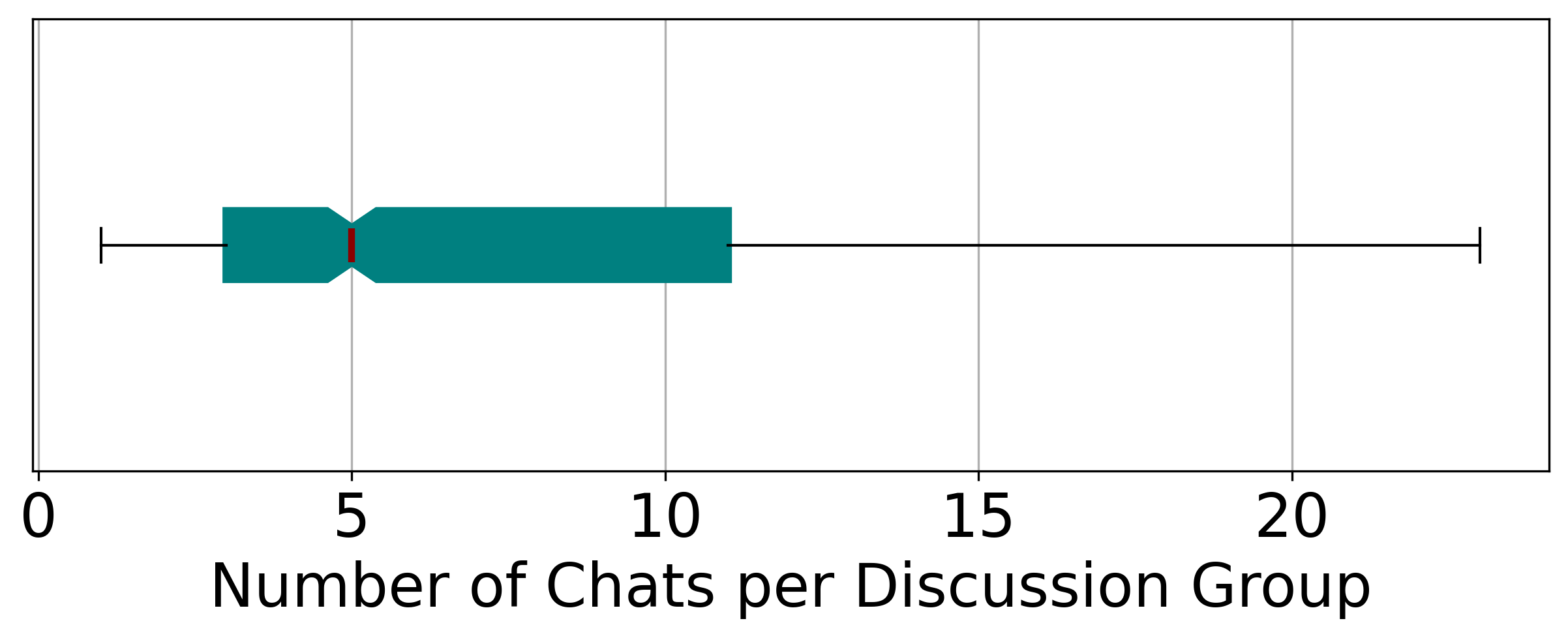}
  \caption{Chat Distribution (outliers truncated). 5.87 chats were sent per group per section on average.}
  \label{fig:chat_dist}
\end{figure}

Figure \ref{fig:chat_dist} shows the student-sent message distribution per group. It is worth noting that the average of 5.87 questions per group per section  (\textasciitilde{}0.84 question per student) only accounts for questions asked through the \textit{Pensieve Discuss} platform. Students were also able to ask questions via the voice channel in Discord during the session, which was not captured from our usage data.

TAs reported that in previous semesters, without using our system, students typically asked 12.5 questions\footnote{Average reported from two TA responses} total in a discussion section of 30 students (0.42 questions per student). This suggests that the adoption of our system substantially increased the question volume.

Since students are working in a group with our system, TAs also testified that students try to resolve their questions within the group first and then reach out to the AI Tutor or a human TA. As such, a small-group environment may reduce the frequency of students over-relying on the AI tutor. 

\subsubsection{TA to AI Tutor interaction}

\begin{table}[h]
\begin{tabular}{|c|c|}
\hline
\textbf{Read}     & \textbf{423} \\ \hline
Endorsed & 188          \\ \hline
Edited   & 3            \\ \hline
\end{tabular}
\caption{614 TA labeled chats out of 7516 chats}
\label{table:chat_label_ta}
\end{table}

Table ~\ref{table:chat_label_ta} shows that approximately 8\% of AI Tutor messages were reviewed by TAs. About 30\% of the reviewed messages were endorsed.

During the interviews, both TAs mentioned that they tried their best to review every message the AI Tutor sent to students, and estimated that they reviewed between 70\% and 90\% of the messages. However, they also acknowledged that they often forgot to mark messages as read, which explains the discrepancy between the review rate in the data and their testimonials.

\subsubsection{TA to Student interaction}

TAs testified that having the TA view in our system helped them understand each group's progress better than in semesters without our system, as they could see the students' code in real-time. 

However, TAs rarely used our system to send chat messages to the students. Out of 1314 sessions, only 61 sessions (4.6\%) included instances where TAs sent messages to the student groups. Until Discussion 9, TAs shared the same chat window with the AI Tutor. This caused confusion, as TAs were unclear whether students were directing their questions to the AI Tutor or to the TA. Additionally, since the AI Tutor responded to every question, even to the follow-up questions intended for the TA, the AI Tutor continued to send messages in the same window as the TA.

To address this issue, we released a separate TA chat window starting from Discussion 9. While TAs preferred the separate chat window, most students continued to use the AI Tutor chat window and rarely asked questions in the TA chat view.

TAs were also able to communicate with students verbally by visiting the group in-person or using Discord. These additional communication methods may have contributed to the low usage of the chat interface for TA-student interactions within our system.


\section{Limitations}

Our study primarily examined the impact of \textit{Pensieve Discuss} on student collaboration by comparing student survey results from different semesters. While the mode of instruction and the small-group tutoring format of discussion sections remained the same, aside from the introduction of our system, there are minor differences in problem descriptions that may have influenced the survey results. 

We also acknowledge that our AI Tutor might be less beneficial to students who are behind in the course content. During TA interviews, one TA mentioned that the system wasn't as helpful for students who were significantly behind in the lecture material, as these students tended to over-rely on the AI Tutor.




\section{Future Work}

\textit{Pensieve Discuss} has the capability to support multiple programming languages. Our CS1 course is taught primarily in Python with other languages, all of which are supported in \textit{Pensieve Discuss}. Instructors can easily select the desired language for each question in our system's Content Management System and upload their content accordingly. 

Additionally, our system is not limited to CS courses. In the future, we aim to leverage the scalability of \textit{Pensieve Discuss} to expand its use to other courses within Computer Science and potentially to other disciplines. 






\section{Conclusion}

We present \textit{Pensieve Discuss}, a software platform designed for scalable small-group CS tutoring. We deployed our platform in a large CS1 course with 800 students at a public university for an entire semester. Through student surveys, usage data analysis, and teaching assistant interviews, we found that \textit{Pensieve Discuss} helps increase student collaboration, student satisfaction, and question volume during small-group tutoring sessions.

\textit{Pensieve Discuss} is a first attempt to integrate LLM to scale and improve small-group tutoring section. We hope that the approach taken by our system serves as a stepping stone for future research in the application of AI in small-group tutoring.

\bibliographystyle{ACM-Reference-Format}
\bibliography{sample-base}

\end{document}